# Emission Reduction in Urban Environments by Replacing Conventional City Buses with Electric Bus Technology: A Case Study of Pakistan

Muhammad Haris Saleem[1*], S. Wajahat Ali[1], Sheikh Abdullah Shehzad

*Lahore University of Management Sciences, DHA phase 5 sector U, Lahore, Punjab, 54792, Pakistan*
*Corresponding author (21100088@lums.edu.pk)

**Highlights:**

- Possibility of transition to Electric Buses given support from government entities.
- Major challenges were establishing infrastructure and national-level policy for Electric Vehicles.
- To catch the public interest, authorities must play a large role in incentivizing E.V. in comparison to incentivizing Diesel.
- Regulations for Advanced charging infrastructure are to be prioritized.

**Abstract:** The global transportation industry has become one of the main contributors to air pollution. Consequently, electric buses and green transportation are gaining popularity as crucial steps to reduce emission concerns. Many developed countries have already adopted the concept of Battery Electric Buses (BEBs), while the developing ones are just starting with it. However, BEB fleets have advantages, such as lower fuel, higher efficiency, lower maintenance, and energy security. Yet, several obstacles must be overcome to support the mass deployment of BEBs. These incorporate forthright expense charges, arranging loads, BEB reach, and newness to BEB innovation. Stakeholders like policymakers, private company owners, and government leaders have a lot to consider before introducing BEBs at any level in Pakistan. As a result, to operate an electric bus system profitably, it is crucial to develop a proper electric bus network and fleet, especially for bus operators who need to buy enough electric buses at the appropriate time. As a result, this paper aims to investigate if operating an electric bus could be an alternative to regular bus operations. The proposed methodology develops modeling software to cater to various scenarios to determine a proper-designed electric bus operating system in terms of the electric bus route, service frequency, and quantity. This research work simulates and financially analyses an operating Public Transport Infrastructure with a proposed Green Solution. The results show that regardless of the high upfront costs of BEB infrastructure, it becomes **profitable** in **6-7 years**, resulting in a decreased **Total Cost of Ownership (TCO)** of approximately **30%** of its counterpart. The study also provides a clear policy pathway to help stakeholders make informed decisions related to the electrification of public transport in Pakistan.

## 1. Introduction

Globally, it was discovered that road transport emissions seriously jeopardize the quality of the air in inner cities and contribute to global warming (due to the emission of Carbon dioxide, $CO_2$). Traffic congestion, air pollution, and noise pollution are three externalities associated with transportation operations that call for the attention of the appropriate authorities (including transport policymakers, operators, the public, etc.). These externalities must be appropriately addressed to ensure the sustainable expansion of transportation, given that the transportation industry accounts for more than **25% of global energy consumption** [1]. Carbon dioxide ($CO_2$), Methane ($CH_4$), and Nitrous Oxide ($N_2O$) are precisely the main contributors to Green House Gas (GHG) emissions from the transportation industry (N2O). Adopting electric buses as a practical mode of public transportation proves to be a successful endeavor to launch green mobility to protect the environment. The environmentally friendly and designed electric bus is an emerging technology to lower carbon emission levels. One of the most promising solutions, the use of electric vehicles, maybe a viable way to alleviate the environmental issue. Figure 1 shows the CO2 Emission trend in Pakistan.

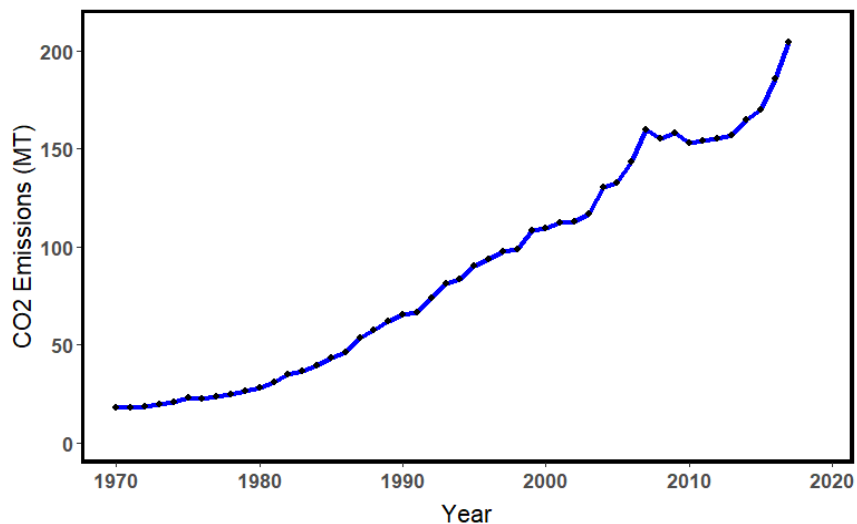

Figure 1: CO2 (M.T.) Emissions in Pakistan [5]

Pakistan contributes 0.9% [2] of the total global greenhouse gas (GHG) emissions, most of which come from the country's transport sector. In 2018, this sector alone contributed 51.3 MT of CO2 eq. and 20-25% on average to the cumulative emissions [2]. Pakistan, being an agricultural country, is facing a climate emergency. It is the 5th most climatically affected country worldwide. Having faced approximately 150 weather incidents resulting from climate change, it was ranked 19th out of 181 on the scale to measure fatality rate. The Asian Development Bank (ADB) and World Bank (W.B.) have published a report on the "Climate Risk Country Profile," which estimates that an increase of up to 2.5 degrees Celsius in temperature over the next two decades will lead to an annual loss of up to $3.8 billion. The publication reports a 44% increase over five years (2013-2018). In 2019 alone, 28% of the 199 MT CO2 emissions from fuel combustion resulted from the transport sector. The country loses 8.84% of its gross domestic product (GDP) annually due to environment-caused diseases [4]. Table 1 below shows different climatic disasters Pakistan has faced over the years and how they have affected the economy and people's lives.

Table 1. Climatic disasters faced by Pakistan.

| Disaster Type | Events Count | Deaths | Total_Affected (Million) | Damage (Million $) |
|---|---|---|---|---|
| Drought | 1 | 143 | 2.2 | 247 |
| Extreme Temperature | 15 | 2759 | 0.08 | 18 |
| Floods | 75 | 11104 | 64.3 | 19911.3 |
| Storm | 18 | 1424 | 2.2 | 1710.9 |

The world is moving towards greener resources for almost everything, transportation or energy. Reducing the carbon footprint is the focus of the world. Transportation is the primary source of emission, including two heads of emission. Emissions from fuel in case of fuel transport, and emissions from the energy mix on which our electric vehicles will operate. Table 2 below shows greenhouse emissions in the transport sector and its forecast based on previous trends.

Table 2. Projected Green House Gas Emissions in the Transport Sector (MT CO2e) [21]

| **Emission Source Category** | **2012** | **2015** | **2020** | **2025** | **2030** |
|---|---|---|---|---|---|
| Road Passenger Vehicles | 21.1 | 23.5 | 29.1 | 36.1 | 45.2 |
| Road Freight Vehicles | 12.6 | 14.5 | 18.9 | 24.2 | 30.6 |
| Aviation | 1.5 | 1.7 | 2.4 | 3.2 | 4.3 |
| Rail | 0.2 | 0.3 | 0.4 | 0.5 | 0.7 |

Pakistan, among other countries, has published a National Electric Vehicle Policy (NEVP) to shift to 50% of new buses sales by 2030 and 90% by 2090 [6], which will significantly reduce the tailpipe emissions from transportation. To mitigate the carbon footprint from the energy mix, Pakistan is shifting to a greener energy resource. Pakistan has adopted an Alternate and Renewable Energy (ARE) policy to go 60% green by 2030. Currently, Pakistan produces 42% Green Energy [7], and by 2030 NEPRA Forecasts this to be 75% Green Energy [8], which will result in 42.5% fewer carbon emissions from our energy mix, reducing from 0.416 Kg/kWh to 0.239 Kg/kWh. Given the forecast, Pakistan is expected to reach its goals within the timeline.

Table 3. Current Energy Mix and IGCEP Forecast of Pakistan

| Energy Mix (Year) | Emissions (kgCO2e/kWh) |
|---|---|
| 2020 | 0.416 |
| 2025 | 0.351 |
| 2030 | 0.239 |

Electric buses look physically like regular gasoline buses, but their operating mechanism is entirely different. The primary differences between the two buses are their various power-producing and handling systems. In technical terms, a typical bus runs on Diesel or gasoline and internal combustion to transform chemical energy into mechanical energy. On the other hand, an electric bus is powered by chemical energy that is converted to electrical energy in a battery box (generated by an electric motor).

The purpose of an electric motor is to move a vehicle by drawing power from a battery module. Additionally, the engine system of a conventional bus contains more moving components than an electric bus, which has a motor. As a result, traditional buses require frequent oil changes, tune-ups for better mileage, exhaust system maintenance, and other maintenance tasks. On the other hand, sealed lead-acid batteries, which require no maintenance, are used in electric buses. Conventional buses contribute to climate change by emitting $CO_2$, whereas electric buses are more environmentally friendly.

However, current electric bus technologies have raised some operational limitations. The limited storage capacity of the batteries has brought up the battery constraint. As a result, the range and duration of an electric bus are constrained. Other attributes of the battery, such as its high price, bulky battery pack, extended charging time, and others, have a significant impact on the use of electric buses. A well-designed electric bus operating system must account for these operational restrictions.

Accordingly, this study examines the feasibility of switching to an electric bus by reviewing the potential advantages of several scenarios (including financial and environmental benefits). The proposed research allows the bus operator to decide how many electric buses to purchase (in fleet planning decisions). In contrast, the developed electric bus network design can determine a desired bus route and service frequency. This is done by considering a variety of operational constraints. Additionally, this study contributes to the field of travel forecasting, notably in demand modeling. Specifically for the research region in Lahore, Punjab, Pakistan, the goal of this work is to evaluate and analyze the performance of various methods (scenario-based), which may end up being the most desirable scenario for operating the electric bus system in a better manner. It is crucial to examine numerous alternative strategies because doing so helps to understand the functions of various tactics and pinpoint the best operating system for electric buses. The scenario-based analysis is critical to identifying the ideal scenario for running an electric bus. Comparing operating scenarios with various functional features will help with this (including routes, headway, and charging facility). The established methodological framework can compare electric bus performance from a supply and demand standpoint by setting up various scenarios. Overall, it is predicted that the suggested methodological framework and the findings would help the bus operator purchase and run electric buses in an environmentally friendly manner, which will also benefit the bus users.

This paper is organized by discussing the state-of-the-art literature review in the Section "Literature review," then the methodological framework is established in the next section named "Methodology. A case study is examined with a thorough discussion to determine the relevance of the suggested study in the "Case Study" section. Section "Conclusions" follows to wrap up this essay.

## 2. Literature Review

This section reviews the literature on electric buses by pointing out the issues hindering Electrical buses' deployment. It further elaborates on the related studies that focus on the performance and operations of electric buses in terms of economic, environmental effects, energy, and operational aspects. In contrast with the Internal Combustion Engine (ICE) operated buses, BEBs are battery-electric buses driven by electric motors. These buses operate on lithium-ion batteries, so they have zero tailpipe emissions, and their operation is also more noise-effective than diesel buses [9]. The primary operational hindrance in the deployment of BEB is

1. Range Anxiety due to limited battery capacity and long charging times
2. Limited charging infrastructures
3. Non-readiness of penetration of BEB into Grids.

The usable energy of these batteries is typically found to be 70% of the theoretical battery capacity at the start of its life. Battery size for BEB, like their size, also varies from 76 kWh to 660 kWh of capacity, giving out a range of up to 450 km depending upon the usage and load on the battery [10]. Regarding the battery capacity, the energy consumption and maximum vehicle range of E.V.s are proportional to the traveled distance, which is considerably influenced by the driving and environmental conditions, e.g., the speed profile and air conditioning requirement. Moreover, it further depends on the battery age, capacity, and load the bus carries. E.V.s need to have a minimum electricity level to travel. As such, the traveled distance and time of E.V.s are relatively limited; hence, E.V.s require frequent charging, which must be performed at a specific charging station. Besides, range anxiety (i.e., the fear of a driver having insufficient electricity to reach the destination) is another concern for using E.V.s. Range anxiety limits consumer adoption and social advantages since E.V. users can be compelled to restrict their driving to small distances. Effectively implementing public charging infrastructure is one method to alleviate range anxiety.

Two techniques are used to provide the charging infrastructure for these buses: (1) Conductive On-Route, and (2) Inductive charging. In the case of conductive, automatic docking or plug-in chargers are used. A.C. and D.C. charging comes under this type of charging, where level 1 and level 2 chargers are considered A.C. and take several hours to charge the bus. On the other hand, level 3 and ultra-fast charging use D.C. power and can charge the bus within 5-20 minutes depending upon supplying power that can vary from 175–500 kW [11]. Inductive charging is known as wireless charging, where coils are buried beneath the bus stops. Fast charging is being developed in such a way that it will be able to recharge an E.V. in less than 5 min [12], but it can adversely affect the battery health by shortening its life.

Conversely, a slow charging station charge requires a long time to charge an Electrical vehicle, generally 2–8 h with Level 1 or 2 (110–240 V) [13]. Therefore, the setting time has significantly influenced the public acceptance of E.V.s. A battery Swapping station that replaces old batteries with new ones is an attractive approach; this technique saves time and reduces the upfront cost of electric vehicles [14]. It requires less than 5 – 10 min, but the only disadvantage is that all vehicles must use identical battery banks. Moreover, a hefty amount of capital and appropriate policies is required to run the swapping station [14]. However, a battery swapping station can increase the uplift of BEB by reducing the reliance on the charging station.

Electric vehicles are highly adapted to the environment and can be an excellent alternative to diesel vehicles so that environmental pollutants may be reduced. These may be air or noise pollutants. The Greenlining Institute issued a report on the benefits of electric buses, according to which electric buses are superior to other buses for the health of the community and the region's climate, benefiting the reduction of respiratory diseases and the removal of strokes for millions of households in the area. According to [15], the amount of particulate matter (PM) and nitrogen oxide (NOx) in electric buses is lower than those of diesel buses, batteries, fuel cells, and compressed natural gas.

BEBs will not only help mitigate the threat to air quality but also help reduce greenhouse emissions. The bar graph below in Figure 2 [22] compares different types of bus technologies and GHG emissions of each kind. Amidst the growing population and increasing number of vehicles on the road, the air quality will only degrade if the emissions go unchecked.

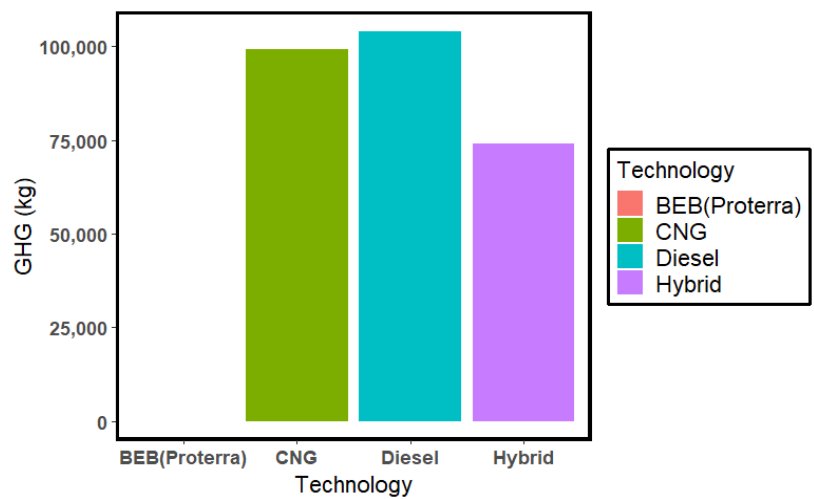

Figure 2: Green House Gas Emissions by Bus Technology

Pakistan is a net importer of petroleum. An increase in oil prices in the international market jolts the value of the country's currency and economy. The rise in oil prices is directly related to inflation. Figure 3 [3] shows the Pakistan oil bill from 2009 to 2020. The transport sector is a significant part of oil consumption in the country, and its share has increased over time, accounting for over 17 million tons. Over 90% of the country's transport relies on imported petroleum. Hence, shifting towards BEBs, which will run on electricity produced using local fuel, renewable resources, and hydro potential, will decrease the oil import bill and strengthen the country's economy by preserving precious foreign reserves.

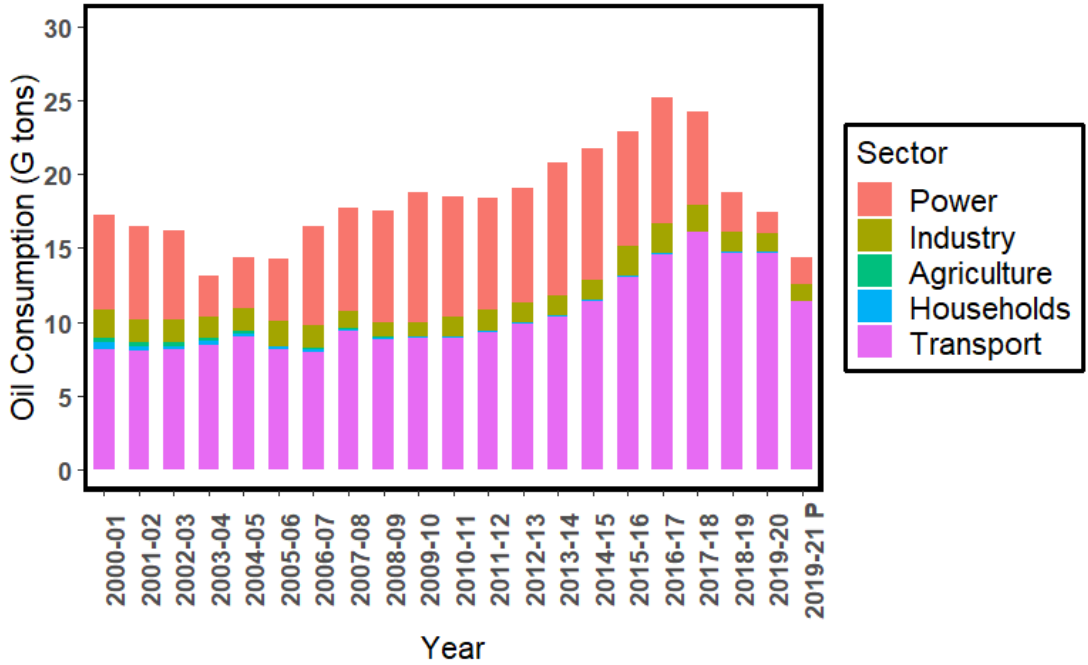

Figure 3: Oil Consumption Trend from different sectors in Pakistan

In terms of Energy Efficiency, electric vehicles are four times more energy-efficient than conventional vehicles converting 77% of Electricity to Power, while traditional cars convert 12-30% of fuel to power [16]. Energy Storage Systems (ESS) are crucial to energy efficiency, frequently measured as the net volume of energy used for one kilometer of travel [17]. Energy density, power density, and cost are the primary factors in assessing ESS efficiency. According to [18], the supercapacitor is the best energy storage technology (capable of quick charging) to address the issue of lengthy charging times and inadequate battery life. The bar graph in figure 4 [22] shows the mpg for different vehicle technology types. It can be observed that BEB outperforms all other technologies.

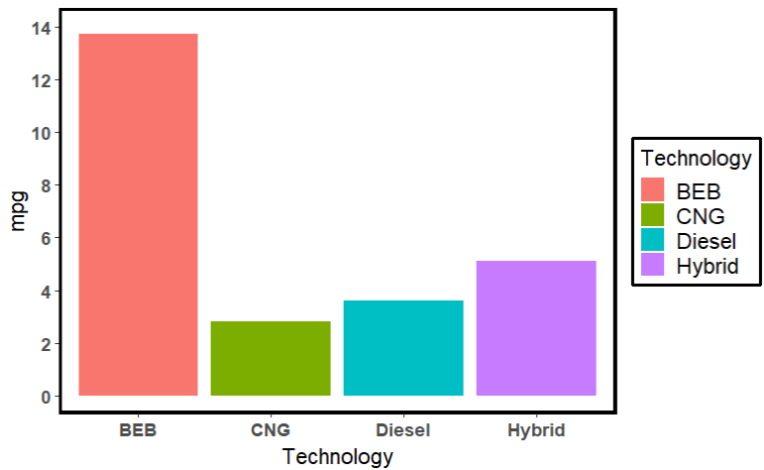

Figure 4: Miles Per Gallon for Different Vehicle Techniques

The proposed methodology in [19] involves four main stages, namely (i) model development, which makes use of EMME (Equilibre Multimodal, Multimodal Equilibrium) modeling software, (ii) data collection & compilation, which is based on the number of vehicles converted to Passenger Car Unit (PCU) for consistency, (iii) traffic and transit system calibration & validation which involves O.D. matrix formation representing trips between centroids. The O.D. matrix is integrated into EMME to run the standard traffic assignment through the modeler. The output in terms of the number of vehicles passing through numerous links, junctions, roundabouts, and interchanges is generated. In the final stage (iv), bus network design & fleet planning determines a desired electric bus route with the corresponding bus frequency and quantity. Hence, the output resulting from the four stages mentioned above includes the electric bus route, frequency, and quantity to serve the passengers better. Concisely, the studies described above showed that a feasible operating system for electric buses requires addressing several elements and concerns. However, limited studies simultaneously analyze electric bus route design and fleet planning (especially scenario-based analysis). Besides, the existing studies mainly focus on the big cities in developed countries. In other words, relevant research for a developing country is exceptionally scarce. Thus, a properly designed case study and its result will be required to analyze the readiness to uptake BEB, particularly for developing and underdeveloped countries.

## 3. Methodology

This article intends to investigate the viability of operating electric buses (to replace the operation of conventional buses) in terms of electric bus network design and fleet planning, in line with replacing fossil fuels as the only energy source for the transportation sector. There are two aspects of this paper. One is the operational feasibility of electric buses, and the second is the financial feasibility of conventional bus technologies available.

Operational feasibility is analyzed by simulating the operation of any available infrastructure. For any public transport, operational efficiency would involve Energy Consumed, Distance Travelled, and the time each passenger must wait for their transportation. This paper will quantify the abovementioned factors and compare them to conventional transport figures.

The Simulation Model is developed in R Studio based on data from relevant authorities' interviews for the individual cases. However, for any Battery Electric Transport, the infrastructure plays a pivotal role. The infrastructure includes Batteries, Chargers of different types and ratings, passengers, Stations, and several Buses. The simulation generates passengers and chargers based on fleet size requirements and total passengers. Passengers developed also vary according to **peak and off-peak** hours in operational timing keeping the limit of total passengers provided. The simulation is then calibrated to minimize bus waiting time at each station. It outputs the energy consumption, distance traveled by each bus, and, most importantly, the waiting time for passengers on average.

For financial feasibility, we will consider all the possible associated costs mentioned in Figures 5 and 6 below and compare the Total Cost of Ownership (TCO) of conventional buses and Electric Buses. In this analysis, the impact of both buses on the environment concerning carbon footprint and tailpipe emission will be mapped to a socio-environment cost and incorporated into our study.

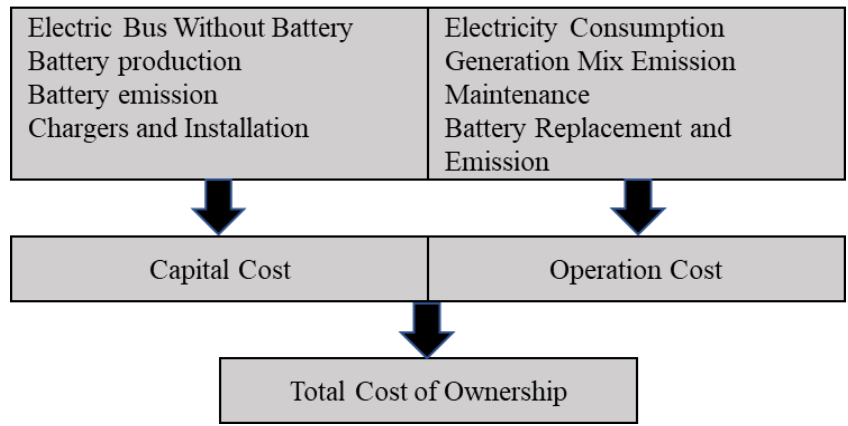

Figure 5: Costs Associated with Battery Electric Bus

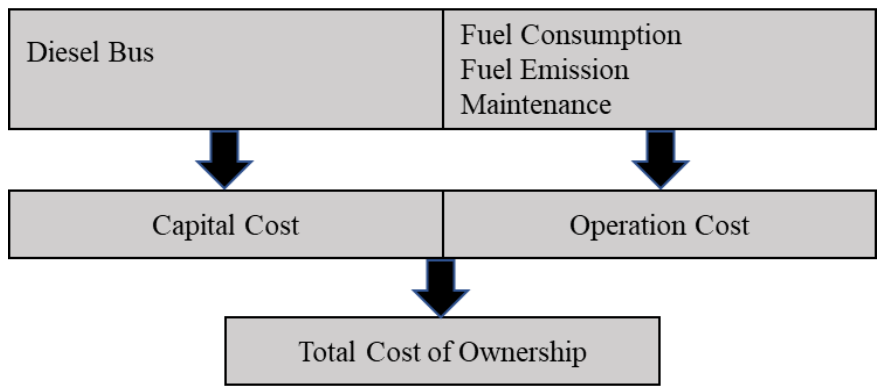

Figure 6: Costs Associated with Diesel Bus

The initial assumption for analysis is that emission cost from the production of the base structure for Diesel and electric buses, including the motors and engine, and associated production cost is assumed to be similar.

For Diesel Buses, engine and fuel type are vital; emissions depend on engine efficiency and fuel quality. In the case of Pakistan, both are Euro 2. The primary emission for diesel buses is from the tailpipe. The diesel carbon emission constant for Euro 2 is 2.6 kgCO2e per liter of fuel [20]. Using equation 1, we calculate the energy consumed. Given fuel consumption and Diesel per liter emission constant, we find the Carbon Dioxide emitted from a single bus annually and can map it to its effect on the environment.

$$FC = DT_{DB} / M_{DB} \qquad (1)$$

Where,
*F.C. = Fuel Consumption in Litres*
*$DT_{DB}$ = Distance Travelled by Diesel Bus*
*$M_{DB}$ = Mileage of Diesel Bus*

For Battery Electric Buses, battery production and energy mix emissions are also considered. For batteries, we also look at the battery replacement cost depending upon the replacement period of Lithium batteries. Battery production emission depends upon battery size, for which we use a standard figure for emissions per kWh and scale it according to our requirements. Similarly, the cost of Chargers (Fast or Slow) also depends on the charger's rating. Hence, the per kWh standard is used and scaled accordingly.

Using the bus mileage and distance traveled shown in equation 2, we find the total units consumed. The energy mix emission constant helps us find the effect of Electric Buses on the environment.

$$EC = DT_{EB} / M_{EB} \qquad (2)$$

Where,

*E.C. = Electricity Consumption in kWh*
*$DT_{EB}$ = Distance Travelled by Electric Bus*
*$M_{EB}$ = Mileage of Electric Bus*

# 4. Case study:

## 4.1 Electrification of Metro Lahore

The Lahore Metrobus Project was first proposed in 1991, and then the construction was started in March 2012. The service first began in February 2013 and is operating to date. The project stretched over 26.1 km on Ferozepur Road, starting from Gajju Mata and ending at Shahdara. There are 27 stations on the route, and a total of 64 buses are present in the fleet.

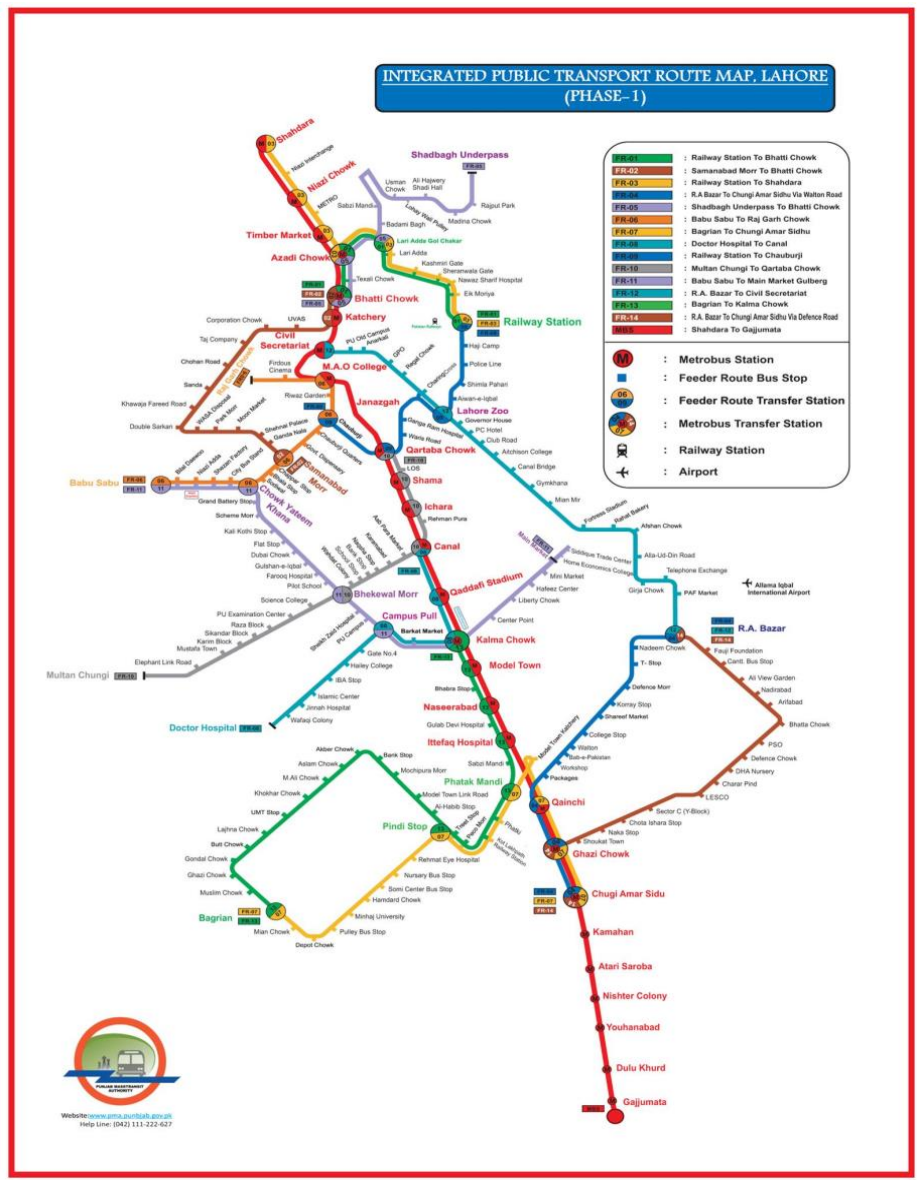

Figure 7: Transit System of Lahore (Red line shows Metro) [23]

The linear nature of the route makes this project more feasible for Electrification due to low infrastructure and planning costs.

Out of Sixty-four, fifty-eight buses are operated daily on the circuit, while six buses are kept on standby in case of peak hours or any emergency. Every bus completes 12 to 13 trips daily, so the total distance traveled by each bus is approximately 320km. The buses used in Metro Lahore are of Diesel Euro 2 technology and, on average, have a mileage of 2km/l.

The bus's average speed is 40 km/h, but it can go up to a maximum of 55 km/h. The average daily passenger served by the metro is 95,000, but the maximum number of passengers in a single day is 175,000. The bus capacity is 160, out of which the seating capacity is 35, and the rest is standing. However according to PMA, they have also boarded a maximum of 220 passengers in a single bus. The buses are operational from 6 15 am to 10 15 pm while there are three peak timings throughout the day, i.e., 7: 00 am to 10: 00 am, 12: 00 pm to 3: 00 pm, and 5: 00 pm to 8: 00 pm. The average time interval between two buses at any station in off-peak hours is 3 minutes, while in peak hours, the time reduces to 2m 15 sec due to the availability of extra buses on the circuit.

Moreover, bus schedules can vary on national holidays and local festivals. Punjab Mass Transit Authority (PMA) leased these buses from a private company at 350 per km. This included everything from fuel to maintenance. PMA has set the bus fare at Rs.30. The Parameters for Lahore Metro are shown in Table 4 below:

Table 4: Lahore Metro Case Study Parameters

| Parameter Name | Lahore Metro |
| --- | --- |
| Number Buses | 64 |
| Bus Capacity (sitting, standing) | 160 (35,125) |
| Mileage Efficiency | 2 km/l |
| Route Length | 26.1 km |
| Average Distance by each Bus | 313.2 km |
| Passengers served per day | 95000 |
| Bus Technology | Diesel Euro 2 |
| Average Speed | 40 km/h |
| Maximum Speed | 55 km/h |
| Maximum Passengers per day | 175,000 |
| Buses in Circuit | 58 |
| Operational Timing | 6:15 am - 10:15 pm |
| Peak Time | 7:00 am - 10:00 am |
| | 12:00 pm - 3:00 pm |
| | 5:00 pm - 8:00 pm |
| Range without refuel | 400 km |
| Time for the bus to arrive in peak hours | 2 mins 15 sec |
| Time for each bus to arrive in off peak hours | 3 mins |
| Cost (Operational, procurement, maintenance) | 350/km |
| Fare per person | Rs. 30 |

A simulation tool is developed to assess bus electrification feasibility for Lahore Metro Bus Service. The primary purpose of this simulation was to find out the number of buses, their battery capacity, chargers, and the amount of electricity required to operate the daily operations and to achieve a similar waiting time for the passengers as the current Metro Lahore operation.

### 4.2. Simulation Model

Figure 7 below shows the flowchart of the metro simulation along with all input and output parameters.

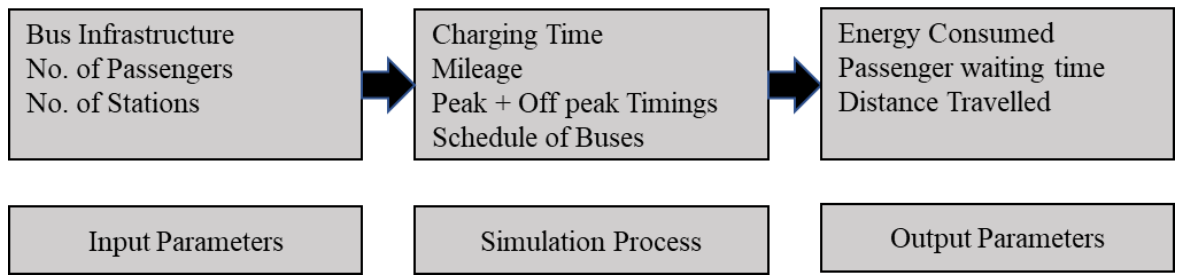

Figure 7: Flow of Metro Simulation

### 4.2.1. Inputs of the Simulation

The simulation is provided with many stations, a sequence of stations, and the distance between the stations. The data for Lahore Metro was obtained from the Punjab Mass Transit Authority (PMA). Then the number of buses to be used in the simulation is provided along with the specifications of the buses:
1. Bus passengers' capacity
2. Bus battery capacity
3. Bus energy consumption
4. The average speed of the bus

The charger's setting follows the bus configuration. Here, some fast and slow chargers and their ratings are configured. Lastly, data for passengers arriving at each station over the operational timing of the system are generated using **Poisson distribution**.

$$f(x) = \frac{\lambda^x}{x!} e^{-\lambda} \ldots (2)$$

*Where :*

**λ**: *Average Rate of occurrence – Different for Peak and Off-Peak*

*x: Total number of occurrences i.e. Passengers Arriving*

The mean number of passengers arriving at each station is separately provided for peak and off-peak hours. According to PMA, on average, Lahore Metro serves **95000 passengers daily**. In our case, the average number of passengers arriving in **peak hours was 10 per minute, and off-peak was 3 per minute**. This configuration gave the total number of passengers using the bus service equal to the average number of passengers served by the system.

Later, the direction a passenger will be heading is determined using a **binomial distribution** with probability provided by the station at which a passenger arrived. The probability is calculated using the formula below

$$Pi = \frac{(Sn - Si - 1)}{(Sn - 1)} \quad \text{...} \quad (3)$$

Where :

$P_i$ : Probability of passenger arriving at $i^{th}$ station

$S_n$: Total number of stations

$S_i$: Current stations

Lastly, the station at which a passenger will depart the bus is determined using a uniform distribution.

**4.2.2. Optimization of Parameters**

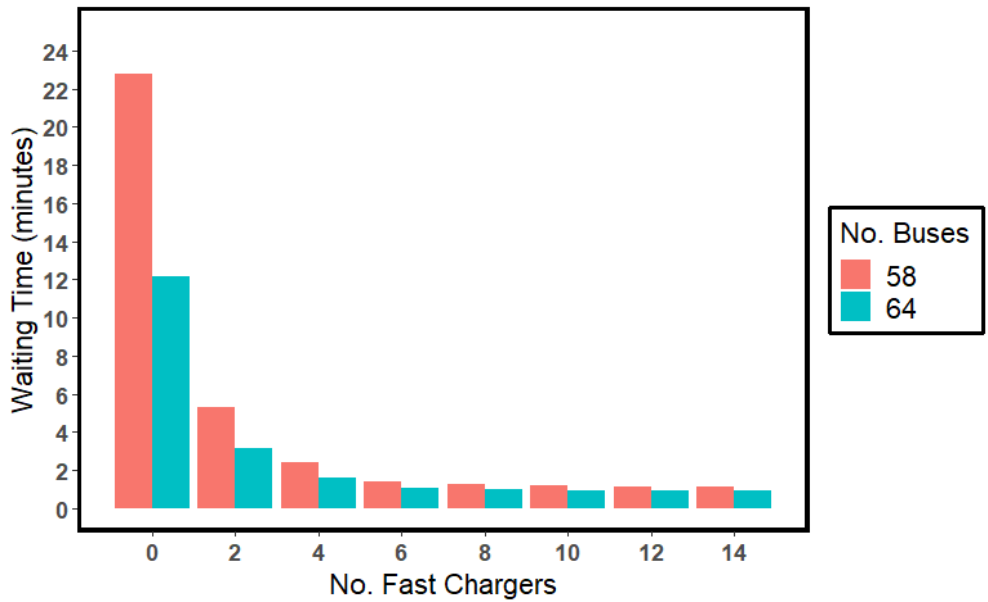

Figure 8: Waiting Time with Fast Chargers for different Numbers of Buses

We have 58 diesel buses on the route in regular metro operation, and 6 buses are in reserve. Assuming the same total number of buses. A comparison of the process with 58 BEB and 64 BEB is made in Figure 8. It can be observed a low waiting time is achieved with the use of 64 buses. In further calculation, No. of Buses will be kept at a constant of 64. The waiting time for 64 buses is **19% less than** 58 buses for 10 fast chargers.

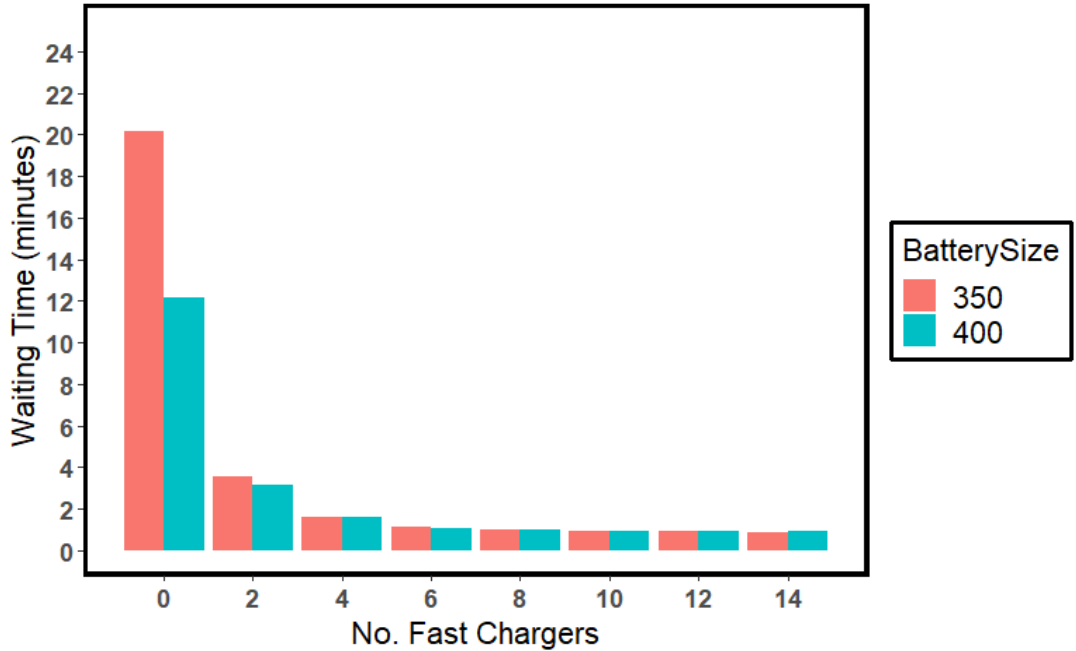

Figure 9: Waiting Time with Fast Chargers for Different Battery Sizes

To optimize the battery size, we analyzed 2 cases, battery sizes of 350 KW and 400 KW. Lower values than 350 KW were not considered because of the C-rating constraint of fast charging. The fast charger we have is of 325KW rating. The results in Figures 9 and 10 show that the waiting period for both cases is almost equal when the number of fast chargers increases, hence the smaller battery size is used to lower the Cost of Ownership.

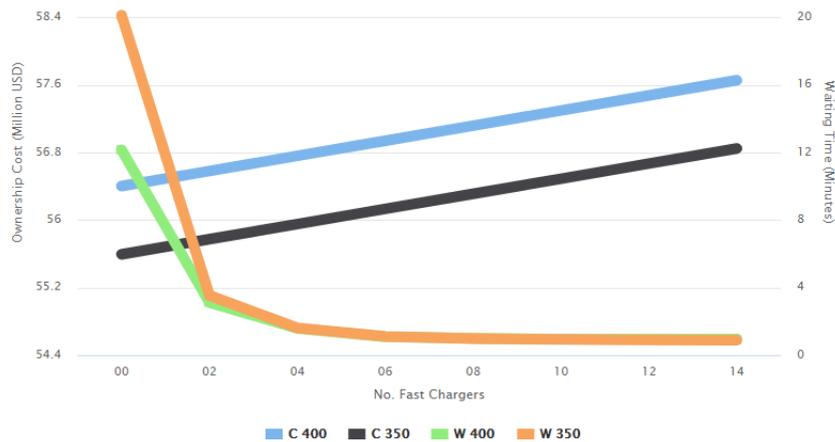

Figure 10: Waiting Time with Fast Chargers for different battery Sizes

The waiting time decreased by 0.04 minutes when we moved from 8 fast chargers to 10. Further increase in the no. of fast chargers has no significance in the decrease of waiting time; however, the cost increased from 8 to 10 Fast chargers by 0.18 million USD, while if we move to 12 chargers, waiting time compared to 8 chargers is decreased by 0.06 minutes. Still, the cost is increased by 0.35 million

USD. We have to see a trade-off between cost and waiting time. At 10 fast chargers waiting time is feasible and hence we do not require more fast chargers.

From the above discussion, our optimized parameters are shown in table 5 below, which will be used to do the detailed financial analysis

Table 5: Optimized Parameters

| Parameter Name | Quantity Options | Optimized Quantity |
|---|---|---|
| Number of Buses | 58, 64 | 64 |
| Battery Size | 350 KW, 400KW | 350 KW |
| Fast Charges | 2,4,6,8,10,12,14 | 10 |

### 4.2.3. Outputs of the Simulation

The following results are derived after the simulation run for an average day operation.

- Average waiting time per passenger
- Electricity required during each minute of the day
- Total distance traveled by each bus
- Energy consumed by each bus
- Total energy consumed by the entire operation

### 4.3. Results and Discussion:

For the designed BEB fleet using optimized parameters, the average waiting time for a passenger was calculated to be 0.9285 minutes, total electricity consumption for a day stood at 38110 kWh where the total distance traveled in a day by all the buses was 23936 km and the average distance by each bus was 374 km. Waiting time compared to regular metro operation is 8% more. However, the cost-benefit of 15% is achieved.

Tables 6 and 7 show, parameters and associated costs with vehicle type for the financial analysis

Table 6. Diesel Bus Parameters and Associated Costs

| Quantity | Price/Units |
|---|---|
| Fleet Size | 58 |
| Bus Cost | £ 350,000 |
| Study Life | 12 Years |
| Fuel Average | 2 km/l |
| Avg. Distance Travelled | 415 km |
| Maintenance Cost | 0.3921 $/km |
| Fuel Emission | 0.002910 Ton/L |
| Emission Cost | 50 $/Ton $CO_2$ |
| Fuel Price | $0.7657/L* |
| **Total Cost of Ownership (TCO)** | **$ 74.7 million** |

Table 7: Battery Electric Bus Parameters and associated costs

| Quantity | Price/Units |
|---|---|
| Fleet Size | 64 |
| Bus Cost | £ 532,000 |
| Slow Chargers | 58 |
| Slow Chargers Rating | 150 KW |
| Slow Chargers Cost | $ 50,000 |
| Slow Chargers Installation | $ 17,050 |
| Fast Chargers | 10 |
| Fast Chargers Rating | 325 KW |
| Fast Chargers Cost | $ 495636 |
| Fast Charger Installation | $ 202811 |
| Battery Size | 350 KW |
| Battery Cost | 137 $/KWh |
| Battery Life | 6 Years |
| Battery Salvage Value | 30 % |
| Study Life | 12 Years |
| Fuel Average | 1.88 km/KWh |
| Avg. Distance Travelled | 374 km |
| Maintenance Cost | 0.206 $/km |
| Fuel Emission | 0.0004586 Ton/L |
| Emission Cost | 50 $/Ton $CO_2$ |
| Electricity Price | $0.1143/KWh* |
| **Total Cost of Ownership (TCO)** | **$ 64.14 million** |

Note: * values in the table are taken as per rate on 16[th] August 2022

**Visualization of Financial Analysis:**

The graph in Figure 11 shows an emission cost comparison over the study life of electric buses. Battery Electric Buses (BEB) have only emissions at the production stage and zero tailpipe emissions when in operation. Meanwhile, Diesel Buses (D.B.) have high emissions in operations, making the cumulative cost close to **6.6 million USD i.e., 83.16% more** than that of electric buses.

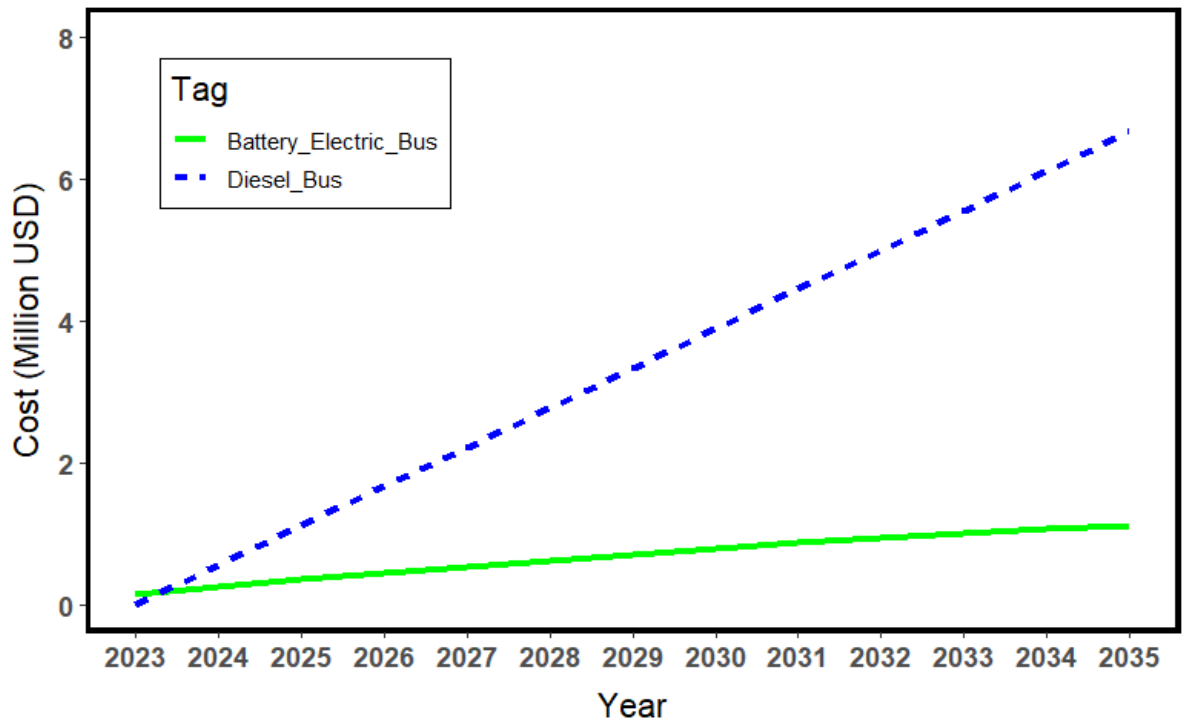

Figure 11: Battery Electric Bus and Diesel Bus Emissions Cost

Figures 12 and 13 show cost projections of diesel and electric buses over the study life. Figure 12 shows the projection when emission cost is not considered, and only energy and maintenance cost is considered in the operation period. While figure 13 shows the cost projection including the emission cost as well. In figure 9, it can be seen if emission is not considered, the breakeven period comes between **2033-2034, while if emission cost is considered breakeven is achieved by 2032**.

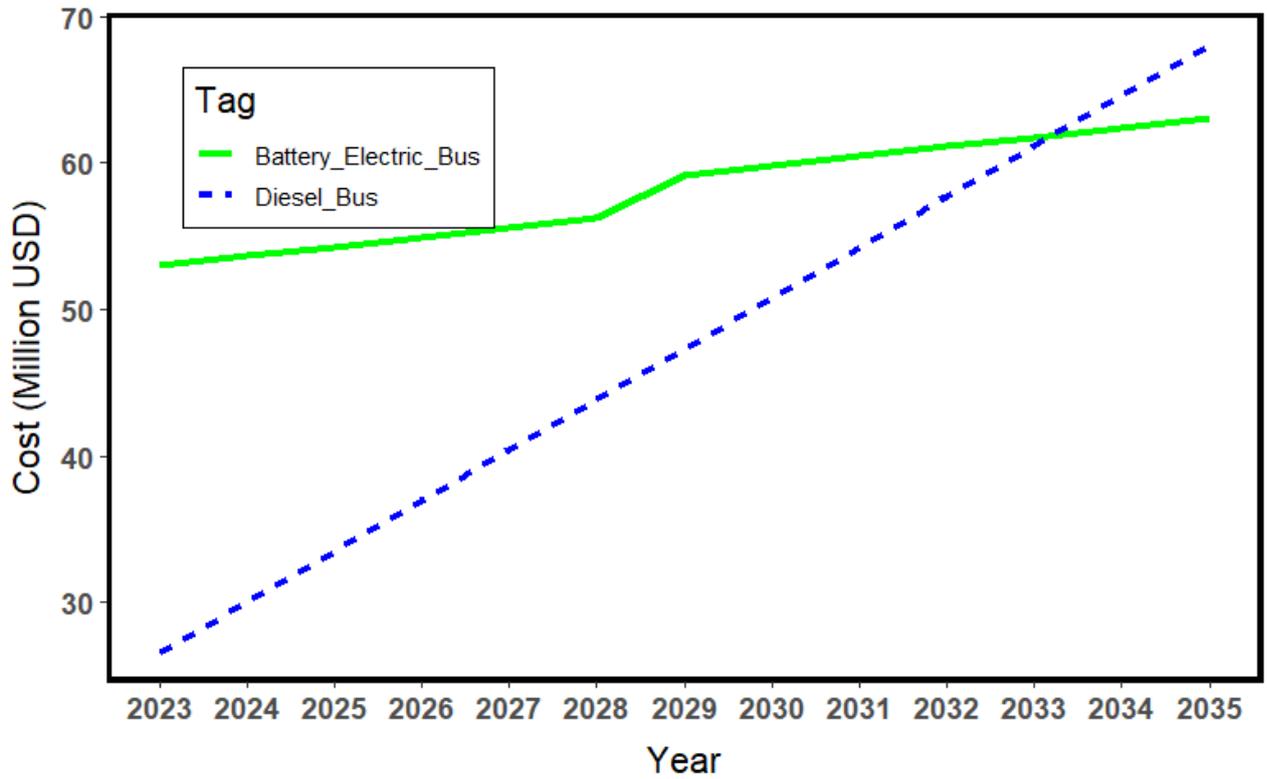

Figure 12: Cost Comparison Battery of Electric Bus and Diesel Bus Without Emission Cost

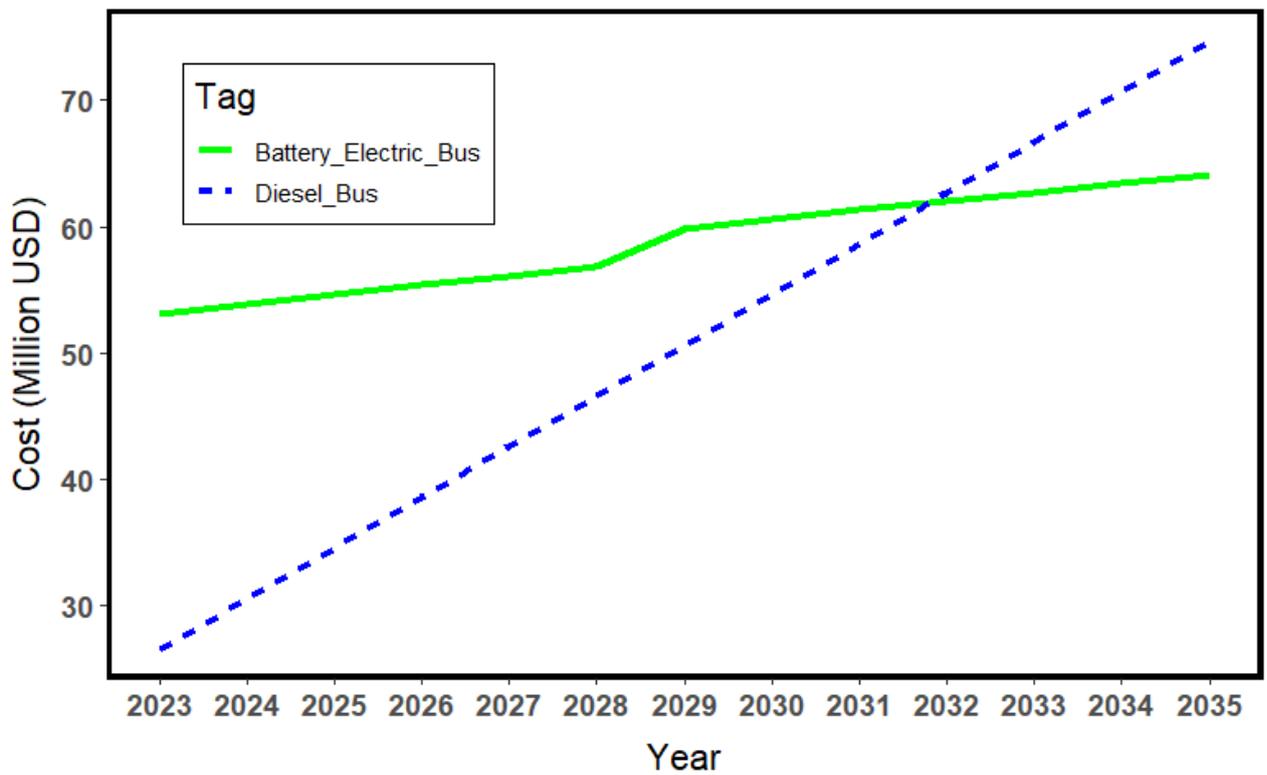

Figure 13: Cost Comparison of Battery Electric Bus and Diesel Bus

The majority component that puts these two technologies apart is energy efficiency. Due to low mileage of diesel buses, a high amount of fuel is used. Fuel prices are following an increase in Pakistan and worldwide.

## 5. Limitations and Policy Recommendations

**Limitations:**

Although, the idea of introducing BEBs seems quite enchanting. However, there are many challenges that need to be overcome for a seamless shift to BEBs. The four different types of challenges faced are

1) Financial:

Although recently the cost of BEBs has been decreasing, BEBs still cost about 1.8 times more than conventional diesel buses. Additional infrastructure costs required for BEBs include charging station costs, installation costs, and coordination with utility companies. In many developed countries such as the United States and Germany, BEB deployments are partially funded by government grants, thus severely impacting the scalability.

2) Planning:

They have different infrastructure constraints, and to compensate for added initial cost, careful planning regarding the routes is required. A few examples of planning scenarios are described below:

    i. Choice of technology: BEBs comes with lots of technology options such as bus size, maximum passenger load, air conditioning, routes, battery size, and capacity.

    ii. Standards and Protocols for Charging Infrastructure: The standards and regulations for the development and installation of charging infrastructure are not well developed.

    iii. Impact on Grid: The installation of the BEB charging network certainly imposes certain requirements on the grid infrastructure.

3) Technological:

Pakistan's average temperature in summers is quite high, and high temperatures pose a potential pitfall for the introduction of electric buses as the degradation of batteries is highly dependent on temperatures. The lack of trained staff to troubleshoot these buses also hinder the feasibility of the project.

4) Operational:

As the deployment of BEBs includes the charging infrastructure, careful planning regarding the routes is required, and the range of BEBs may be constrained by it. BEBs can be charged with Level 1 or Level 2 chargers. When charging on Level 2 chargers which are most commonly available, an electric vehicle takes more than several hours to charge.

**Policy Recommendations:**

1. Ensure policy implementation through a robust regulatory framework to support electric vehicle deployment. While the top-level policy is in place, reflecting high-level aspirations, there has been limited action from relevant agencies in setting out clear directives 9 for taxation, import clearance, instructions, and procedures for implementation of this national E.V. policy. The absence of operational-level work is a fundamental barrier to E.V. market development. For effective policy implementation, key entities need to develop and notify relevant regulations and procedures.
2. Promotion strategy for 'Make in Pakistan' Transition to electric mobility for Pakistan would not be sustainable unless we make efforts to promote the strategy for 'Make in Pakistan.' This requires the development of an indigenous market and channelizing investments into local manufacturing of critical parts, including batteries, motors, and electronics for E.V.s. The most critical part of this entire value chain is the lithium-ion battery. Pakistan would have to make strategic partnerships to secure the supply of key raw materials like Lithium, Cobalt, and Nickel for local production to meet the increasing demand by 2030. China dominates the battery production industry, and we can leverage our partnership with China for local manufacturing of lithium-ion cells in the country.
3. The regulatory framework for Charging infrastructure is critical to streamlining the development of uniform E.V. charging stations across the country. Relevant entities, including NEECA, NEPRA, DISCOs, and PSQCA, shall define clear permitting, licensing, and approval procedures for setting up E.V. charging stations. National Energy Efficiency and Conservation Authority (NEECA) is well-positioned to be designated for facilitating one window operation for setting up, permitting, ensuring compliance, and oversight towards the development and standardization of E.V. charging infrastructure in the country. This would be an essential step in developing the set of technical standards and safety precautions that govern the E.V. chargers to promote and facilitate the sustainable uptake of E.V. charging infrastructure.
4. Establish an optimized Management system for the availability of chargers throughout the day.
5. Establish national fuel economy and emission standards. Despite the fact that the road transport sector heavily relies on oil imports, Pakistan has not yet established the national fuel economy standards for local automotive manufacturers. It is extremely critical for Pakistan to establish its national fuel economy standards for auto manufacturers with a vision to reduce the energy intensity of the road transport sector, depress oil imports, increase energy security and help protect the environment. In this context, the role of NEECA and the Engineering Development Board (EDB) remains critical for the sustainable transition towards clean and efficient mobility in the country.
6. Charging infrastructure plays a key role in enabling and supporting E.V. adoption. DISCOs need to plan ahead for system up-gradation to mitigate the grid impacts – transformers, distribution lines, and switch gears. NEPRA may introduce and employ Time of Use pricing models (off-peak and peak electricity rate) to manage the impact of increasing demand on local distribution networks. Standardization of charging infrastructure is the key to helping develop a safe, reliable, accessible, and affordable E.V. charging ecosystem.

7. Develop adequate financing and technical and human resources capacity to scale up the uptake of E.V.s Currently. There is insufficient expertise on the job market and inadequate technical support to electric vehicle operators and consumers. This necessitates additional training and skill development for practicing mechanics and technicians in workshops and garages to handle EV-related maintenance and operations. Hence, it is essential to introduce capacity-building programs with donors' support for both private and public sector professionals, technicians and engineers.

## 6. Conclusion

As for the effects on the climatic condition of Pakistan, it is evident that deterioration has been fast. The major role-players are GHGs. To observe the contribution of the transportation sector, sources like the World Bank and the Asian Development Bank provide reliable data to support the study being conducted. Through analysis of prior research, it can be concluded that transport emissions per capita are 0.27 T which can be reduced, if the opportunities that Pakistan's National Electric Vehicle Policy provides, are exploited to their full potential.

## Credit authorship contribution statement

**S Wajahat Ali:** Conceptualization, Methodology, Operation Simulation, Writing -original draft, Supervision. **Muhammad Haris Saleem:** Methodology, Financial Analysis, Writing- original draft, Writing – review and editing. **Sheikh Abdullah Shehzad:** Interviews and Data Gathering, Writing- original draft.

**Special thanks to Dr. Kiran Siraj** for providing guidelines and **Dr. Naveed Arshad: for** Supervision and project administration.

# References:

[1] *International Energy Outlook - U.S. Energy Information Administration (EIA)*. International Energy Outlook 2021 - U.S. Energy Information Administration (EIA). (n.d.). Retrieved August 12, 2022, from https://www.eia.gov/outlooks/ieo/

[2] Government of *Pakistan: Updated Nationally Determined Contributors 2021*. Unfccc.int. Retrieved August 12, 2022, from https://unfccc.int/NDCREG

[3] https://www.sbp.org.pk/departments/stats/PakEconomy_HandBook/Chap-2.3.pdf

[4] "Pakistan: Flood Impact Assessment." n.d. http://www.finance.gov.pk/survey/chapter_12/SplSection.pdf.

[5] *Pakistan CO2 emissions, 1970-2021*. Knoema. (n.d.). Retrieved August 12, 2022, from https://knoema.com/atlas/Pakistan/CO2-emissions

[6] Uddin, Moaz. "Pakistan's National Electric Vehicle Policy: Charging towards the Future." International Council on Clean Transportation, November 24, 2021. https://theicct.org/pakistans-national-electric-vehicle-policy-charging-towards-the-future/.

[7] *Power policy - AEDB*. (n.d.). Retrieved August 12, 2022, from https://www.aedb.org/images/ARE_Policy_2019_AEDB.pdf

[8] "Indicative Generation Capacity Expansion Plan." National Transmission and Dispatch Company, Pakistan, May 2021. https://nepra.org.pk/Admission%20Notices/2021/06%20June/IGCEP%202021.pdf.

[9] "Elsevier Enhanced Reader." n.d. Reader.elsevier.com. https://reader.elsevier.com/reader/sd/pii/S1364032116301290?token=51A196531E05E179EE0B7F17F0BAFC7FA15E996BFCF44662B3C8D8F2E8DBE1FD6E44995BE01DEB44049BADFEECE43A7D&originRegion=eu-west-1&originCreation=20220628074657.

[10] Aamodt, Alana, Karlynn Cory, and Kamyria Coney. 2021. "A Product of the USAID-NREL Partnership ELECTRIFYING TRANSIT: A GUIDEBOOK for IMPLEMENTING BATTERY ELECTRIC BUSES." https://www.nrel.gov/docs/fy21osti/76932.pdf.

[11] NASEM. 2018. Battery Electric Buses—State of the Practice. Washington, D.C.: The National Academies Press. https://www.nap.edu/download/25061.

[12] Service, Purdue News. n.d. "Electric Vehicles Could Fully Recharge in under 5 Minutes with New Charging Station Cable Design." Www.purdue.edu. https://www.purdue.edu/newsroom/releases/2021/Q4/electric-vehicles-could-fully-recharge-in-under-5-minutes-with-new-charging-station-cable-design.html.

[13] "The Difference between Level 1 & 2 E.V. Chargers?" 2021. EvoCharge. February 22, 2021. https://evocharge.com/resources/the-difference-between-level-1-2-ev-chargers/.

[14] Chudy, Aleksander. 2021. "BATTERY SWAPPING STATIONS for ELECTRIC VEHICLES." Informatyka, Automatyka, Pomiary W Gospodarce I Ochronie Środowiska 11 (2): 36–39. https://doi.org/10.35784/iapgos.2654.

[15] Syed, Alishbah, Jiquan Zhang, Md Moniruzzaman, Iman Rousta, Talha Omer, Guo Ying, and Haraldur Olafsson. 2021. "Situation of Urban Mobility in Pakistan: Before, During, and after the COVID-19 Lockdown with Climatic Risk Perceptions." *Atmosphere* 12 (9): 1190. https://doi.org/10.3390/atmos12091190.

[16] "All-Electric Vehicles." www.fueleconomy.gov - the official government source for fuel economy information. Accessed February 10, 2022. https://www.fueleconomy.gov/feg/evtech.shtml#:~:text=EVs%20have%20several%20advantages%20over,to%20power%20at%20the%20wheels.

[17] Noman, Syed Muhammad, Afzal Ahmed, and Mir Shabbar Ali. 2020. "Comparative Analysis of Public Transport Modes Available in Karachi, Pakistan." *S.N. Applied Sciences* 2 (5). https://doi.org/10.1007/s42452-020-2678-3.

[18] Riaz, Amna, Mahidur R. Sarker, Mohamad Hanif Md Saad, and Ramizi Mohamed. 2021. "Review on Comparison of Different Energy Storage Technologies Used in Micro-Energy Harvesting, WSNs, Low-Cost Microelectronic Devices: Challenges and Recommendations." Sensors 21 (15): 5041. https://doi.org/10.3390/s21155041.

[19] Teoh, Lay Eng, Hooi Ling Khoo, Siew Yoke Goh, and Lai Mun Chong. 2018. "Scenario-Based Electric Bus Operation: A Case Study of Putrajaya, Malaysia." International Journal of

Transportation Science and Technology 7 (1): 10–25. https://doi.org/10.1016/j.ijtst.2017.09.002.

[20] *How to calculate the CO2 emission from the fuel consumption?* Ecoscore. (n.d.). Retrieved August 12, 2022, from https://ecoscore.be/en/info/ecoscore/co2

[21] *Greenhouse gas mitigation options for pakistan: Transport sector*. (n.d.). Retrieved August 12, 2022, from http://ccrd.edu.pk/files/Transport_LCS%20Factsheet.pdf

[22] *Fuel economy*. Proterra. (2022, April 5). Retrieved August 12, 2022, from https://www.proterra.com/vehicles/zx5-electric-bus/fuel-economy/

[23] "Daewoo Pakistan Express Bus Service | Daewoo Pakistan Express Bus Service." n.d. Daewoo.com.pk. Accessed July 8, 2024. https://daewoo.com.pk/home/lahore-feeder-route-services.